\documentstyle[pra,twocolumn,aps]{revtex}

\begin{document}
\bibliographystyle{prsty}
\wideabs{
\title{High-accuracy wavemeter based on a stabilized diode laser }
\author{Ayan Banerjee, Umakant D. Rapol, Ajay Wasan, and Vasant 
Natarajan\cite{email}}
\address{Department of Physics, Indian Institute of Science, 
Bangalore 560 012, INDIA}

\maketitle
\begin{abstract}
We have built a high-accuracy wavelength meter for tunable lasers using 
a scanning Michelson interferometer and a reference laser of known 
wavelength. The reference laser is a frequency stabilized diode laser 
locked to an atomic transition in Rb. The wavemeter has a statistical 
error per measurement of 5 parts in $10^7$ which can be reduced 
considerably by averaging. Using a second stabilized diode laser, we 
have verified that systematic errors are below 4 parts in $10^8$. 
\end{abstract}
}

The use of diode lasers has become very common in the fields of optics 
and atomic physics \cite{WIH91}. The principal advantages of these 
lasers lie in their low cost, narrow spectral width, tunability over 
several nm, efficient power consumption, reliability, and ease of 
operation. Single transverse mode laser diodes
are available over most of the near-infrared spectrum 
from 600--2000 nm. Furthermore, by placing the diode in an external 
cavity and using optical feedback from an angle-tuned grating, they can 
be made to operate at single frequency (single longitudinal mode) with 
linewidths of order 1 MHz \cite{MSW92,RWE95}. Such frequency-stabilized 
diode lasers have increased the access of researchers to experiments 
which have previously required expensive ring-cavity dye or Ti-sapphire 
lasers. In particular, the field of laser cooling has blossomed in the 
past decade as several alkali atoms and alkali-like ions have cooling 
transitions in the infrared which are accessible with diode lasers 
\cite{nobel97}. However, to use lasers in these applications effectively 
it is important to know their absolute wavelength accurately. In many 
cases, this is achieved by locking the laser to a known atomic or 
molecular transition in a vapor cell. But this may not always be 
possible, especially when working with short-lived radioactive species 
or trapped ions. One solution is to {\it measure} the wavelength of the 
laser precisely by beating its frequency against that of a reference 
laser in a scanning optical interferometer.

In this Letter, we present a design for such a wavemeter in which the 
reference laser is a frequency-stabilized diode laser locked to an 
atomic transition. This gives the reference laser the desired frequency 
stability of 1 MHz and the required wavelength calibration for absolute 
measurements. While expensive commercial wavemeters that use a
stabilized He-Ne laser as the reference are available, our wavemeter
is built in-house around a low-cost laser diode and 
the entire instrument has a component cost of less than \$1500. To 
characterize the accuracy of the instrument, we have measured the 
wavelength of a second stabilized diode laser system. The results 
indicate an accuracy of $5\times10^{-7}$ with a 25 cm scan 
distance, and that systematic errors are below $4\times 10^{-8}$. 

The wavemeter, shown schematically in Fig.\ \ref{schematic}, is based on 
a design first reported by Hall and Lee \cite{HAL76}. The basic idea is 
to obtain the wavelength of an unknown laser in terms of the wavelength 
of the reference laser using a scanning Michelson interferometer where 
both lasers traverse essentially the same path. As the interferometer is 
scanned, the interference pattern on the detectors goes alternately 
through bright and dark fringes. Since both lasers traverse the same 
path, the ratio of the number of fringes counted after scanning through 
a certain distance is the ratio of the two wavelengths. The ratio 
obtained is a wavelength ratio in air, however, the wavelength ratio in 
vacuum (or equivalent frequency ratio) is easily calculated by making a 
small correction for the dispersion of air \cite{EDL66} between the two 
wavelengths. Thus, if the absolute wavelength of the reference laser is 
known, the wavelength of the unknown laser can be determined.

The interferometer consists of a beamsplitter, two end mirrors, and two 
retro-reflectors mounted back-to-back on a movable cart. The 
retro-reflectors, made of three mutually orthogonal mirrors, ensure that 
the return beams are displaced from the input beams and do not feed back 
into the laser. The movable cart is a brass block riding on a 
pressurized air-bearing which allows it to slide freely over a distance
of about 100 cm. The push-pull 
design of the interferometer, with the two retro-reflectors mounted 
back-to-back, has the advantage that the interference pattern goes 
through a complete fringe cycle for every $\lambda/4$ displacement of 
the cart, a factor of two improvement over designs where only one arm of 
the interferometer is scanned. The interferometer produces two output 
beams; of these the one on the opposite side of the beamsplitter from 
the input beam has near-perfect contrast ($>90\%$ in our case) 
because it is a combination of beams 
each of which is once reflected and once transmitted through the 
beamsplitter \cite{foot1}. This beam is detected for both the reference 
laser and the unknown laser by Si PIN photodiodes and the two signals 
fed into a frequency counter. The counter has a ratio function and 
directly displays the ratio of the two inputs with user-settable 
integration times of up to 10 s.

The diode laser system used as the reference laser is built around a 
commercial single-mode laser diode (Mitsubishi ML60125R-01) with a 
nominal operating wavelength of 785 nm and cw output power of 30 mW. The 
light is collimated using a 4.5 mm, 0.55 NA aspheric lens. 
The laser is frequency 
stabilized in a standard external cavity design (Littrow configuration) 
\cite{MSW92} using optical feedback from an 1800 lines/mm diffraction 
grating mounted on a piezoelectric transducer (see inset of Fig.\ 
\ref{schematic}). Using a combination of temperature and current 
control, the diode is tuned close to the 780 nm $D_2$ line in atomic Rb 
($5S_{1/2} \leftrightarrow 5P_{3/2}$ transition). 
A part of the output beam is tapped for 
Doppler-free saturated-absorption spectroscopy in a Rb vapor cell 
\cite{MSW92}. The various hyperfine transitions in the two common 
isotopes of Rb ($^{85}$Rb and $^{87}$Rb) are clearly resolved, as shown 
in the inset of Fig.\ \ref{schematic}. The linewidth of the hyperfine peaks
is 15--20 MHz; this is somewhat larger than the 6.1 MHz natural linewidth
and is primarily limited by power broadening due to the 
pump beam \cite{foot4}. The injection current into the laser diode is 
modulated slightly to obtain an error signal and the 
laser is locked to the $F'=(3,4)$ crossover resonance in $^{85}$Rb, {\it 
i.e.\ }60 MHz below the $F=3 \leftrightarrow F'=4$ transition. From the 
Rb energy level tables \cite{MOO71} and measured hyperfine shifts 
\cite{AIV77}, this corresponds to a frequency of $3.8422958 \times 
10^{14}$ Hz. The elliptic laser beam 
(5.8 mm$\times$1.8 mm $1/e^2$ dia) is directly fed into the interferometer. 
The large Rayleigh ranges ($\sim$34 m and $\sim$3 m, respectively) 
ensure that the beam remains collimated over the length of the 
interferometer.

To characterize the wavemeter, we have measured the wavelength of a 
second identical diode laser system but which was locked to a different 
hyperfine transition in $^{85}$Rb. The measurement serves two purposes: 
first, the scatter in the data gives an estimate of the statistical 
error associated with our instrument since both the reference laser and 
the unknown laser are stabilized to linewidths below 1 MHz (about 3 
parts in $10^9$), and second, the data tells us if there are any 
systematic errors associated with our instrument because the difference 
between the two laser frequencies is already known very precisely 
\cite{AIV77}. The measurements were done with the second laser locked to 
a hyperfine transition in $^{85}$Rb that was 2944 MHz higher than the 
frequency of the reference laser. This implies that the measured ratio 
should be 1.00000766. The actual values obtained are shown in Fig.\ 
\ref{rbhist} as a histogram plot. The data have a fairly good Gaussian 
distribution and the fit yields a mean value of 1.00000763(4) and a 
spread of $5.3 \times 10^{-7}$.

Statistical errors in the measured data arise mainly because the 
frequency counter only detects zero-crossings and does not count 
fractional fringes. The total number of fringes counted depends on the 
fringe rate (or cart speed) coupled with the 10 s integration time. 
Currently our photodiode electronics limits the cart speed so that we 
can use only about 25 cm of cart travel per measurement. This results in 
the single shot statistical error of 5 parts in $10^7$ in the data 
\cite{foot2}. However, the mean value has an error of only $4 \times 
10^{-8}$ since it is an average of more than 100 individual 
measurements. With some improvements in the counting electronics, it 
should be possible to use 50--100 cm of cart travel for each measurement 
and thereby reduce the statistical error per measurement below $1 \times 
10^{-7}$. Resolution enhancement is also possible by phase-locking to
the fringe signal, as described in Ref. \cite{HAL76}. This allows 
fractional fringes to be counted accurately and the statistical error 
to be reduced for the same cart travel. Finally, data contamination
can occur if partial fringes are counted when the cart turns around
at the end of its travel. We therefore take data only when the cart is
travelling in one direction.

It is important to eliminate all sources of systematic error when aiming 
for such high accuracy. The chief cause of systematic error is 
non-parallelism of the two beams in the interferomter. Any misalignment 
would cause an increase in the measured ratio given by $1 / \cos 
\theta$, where $\theta$ is the angle between the beams. We have tried to 
minimize this error by using the unused output beam of the reference 
laser (the one on the same side of the beamsplitter as the input beam) 
as a tracer for aligning the unknown laser beam, and checking for 
parallelism over a distance of about 2 m \cite{foot3}. The consistency 
of the mean value of 1.00000763(3) with the expected value (1.00000766) 
shows that this method works quite well for accuracies up to $4 \times 
10^{-8}$. We have also found it useful to check for parallelism by 
looking for a minimum in the measured ratio as the angle of the unknown 
beam is varied. This works because the measured value is always larger 
than the correct value, whether $\theta$ is positive or negative, and 
becomes minimum when $\theta = 0$.

In conclusion, we have built a high-accuracy wavemeter 
using a scanning Michelson interferometer and a reference 
diode laser. The frequency-stabilized diode laser is locked to an atomic 
transition in Rb for absolute calibration. Using a second 
stabilized diode laser,
we have shown that the statistical error in each 
measurement is 5 parts in $10^7$, and the systematic error is less than 
4 parts in $10^8$. We have recently used this wavemeter to measure the 
wavelength of a diode laser tuned to the 795 nm $D_1$ line in Rb 
($5S_{1/2} \leftrightarrow 5P_{1/2}$ transition) \cite{BRN01}, which 
yields a precise value for the fine-structure interval in the $5P$ state 
of Rb. The precision obtained is an order of magnitude better than 
published values \cite{MOO71} and demonstrates the power of the instrument 
for precision spectroscopy experiments. Hyperfine splittings, which are of
order GHz, are more accessible to techniques
such as microwave resonance or heterodyne measurements \cite{YSJ96}. On the
other hand, fine-structure splittings are of order THz and our technique
is uniquely suited for precise measurements in this range of frequency 
differences.

We are grateful to S.\ Navaneetha for machining the mechanical components 
of the wavemeter. This work was supported by research grants from the 
Board of Research in Nuclear Sciences (DAE), and the Department of 
Science \& Technology, Government of India.

\begin{figure}
\caption{
Schematic of the wavemeter. The wavemeter is a scanning Michelson 
interferometer consisting of a beamsplitter (BS), two end mirrors (M), 
and two retro-reflectors (R). The retro-reflectors are mounted 
back-to-back on a movable cart. The inset on the bottom left 
shows the assembly of the reference diode laser with collimating lens 
and piezo-mounted grating. The inset on the right is a Doppler-free 
saturated-absorption spectrum in $^{85}$Rb as the laser is scanned 
across hyperfine transitions starting from the $F=3$ level. The laser is 
locked to the $(3,4)$ crossover peak corresponding to a frequency of 
$3.8422958 \times 10^{14}$ Hz.
}
\label{schematic}
\end{figure}

\begin{figure}
\caption{
Histogram of measured ratios. The graph is a histogram of the ratios 
measured with the second laser locked to a hyperfine transition of 
$^{85}$Rb that is 2944 MHz away from the reference laser. The solid line 
is a Gaussian 
fit to the histogram, which yields a mean value of 1.00000763(4) and a 
spread of $5.3 \times 10^{-7}$. The mean corresponds to a frequency 
difference of 2932(16) MHz, showing that any systematic errors are below 
16 MHz.
}
\label{rbhist}
\end{figure}


\begin{references}
\bibitem{foot1}
We obtain better than 90\% fringe contrast even though our 
beamsplitter is close to 60/40. 
The other output beam has perfect contrast only if the beamsplitter is 
exactly 50/50. However, as discussed later, this beam is useful in
reducing systematic errors by ensuring that the unknown beam is parallel 
to the reference.
\bibitem{foot4}
Collisional broadening at the gas pressure in the Rb vapor cell 
is estimated to be $\sim$10 kHz and the linewidth of the laser has been 
independently measured to be $\sim$1 MHz.
\bibitem{foot2}
The calculated error from missing one fringe over a distance of 25 cm is 
about $8 \times 10^{-7}$ at 780 nm. The measured error is smaller 
because we are measuring a wavelength ratio very close to 1, and chances 
are high that either both counters see a zero-crossing or miss it. In 
fact, if we try to measure a ratio of exactly 1 by locking both lasers 
to the same transition, we get a bimodal distribution in the measured 
ratios: 1.0000000 most of the time and occasionally 1.0000008 which 
corresponds to one extra fringe counted by the second detector. That is 
why we have chosen to measure a test ratio of 1.00000766 where the 
distribution is Gaussian.
\bibitem{foot3}
Ideally, the unknown beam should enter exactly where the reference beam 
exits so that both beams traverse exactly the same path 
in opposite directions. However, this causes one laser to feed into the 
other and destabilize it. Therefore we align the beams with a small 
offset and check for parallelism over a finite distance.
\end{references}
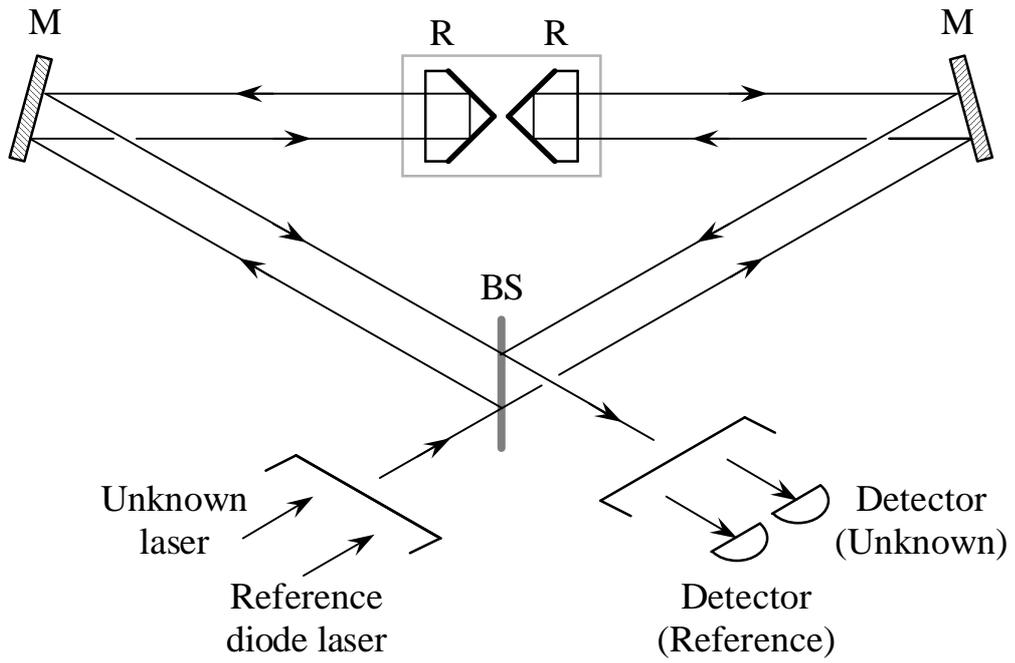
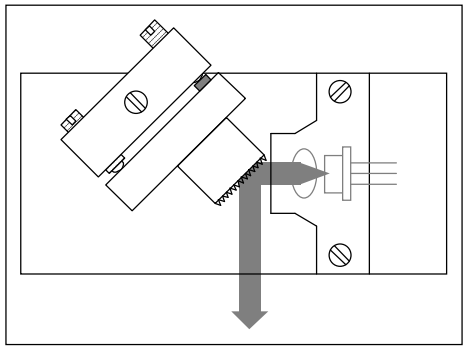
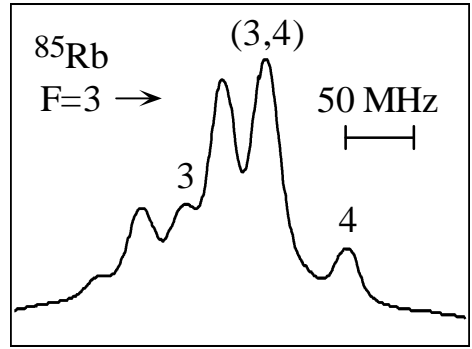
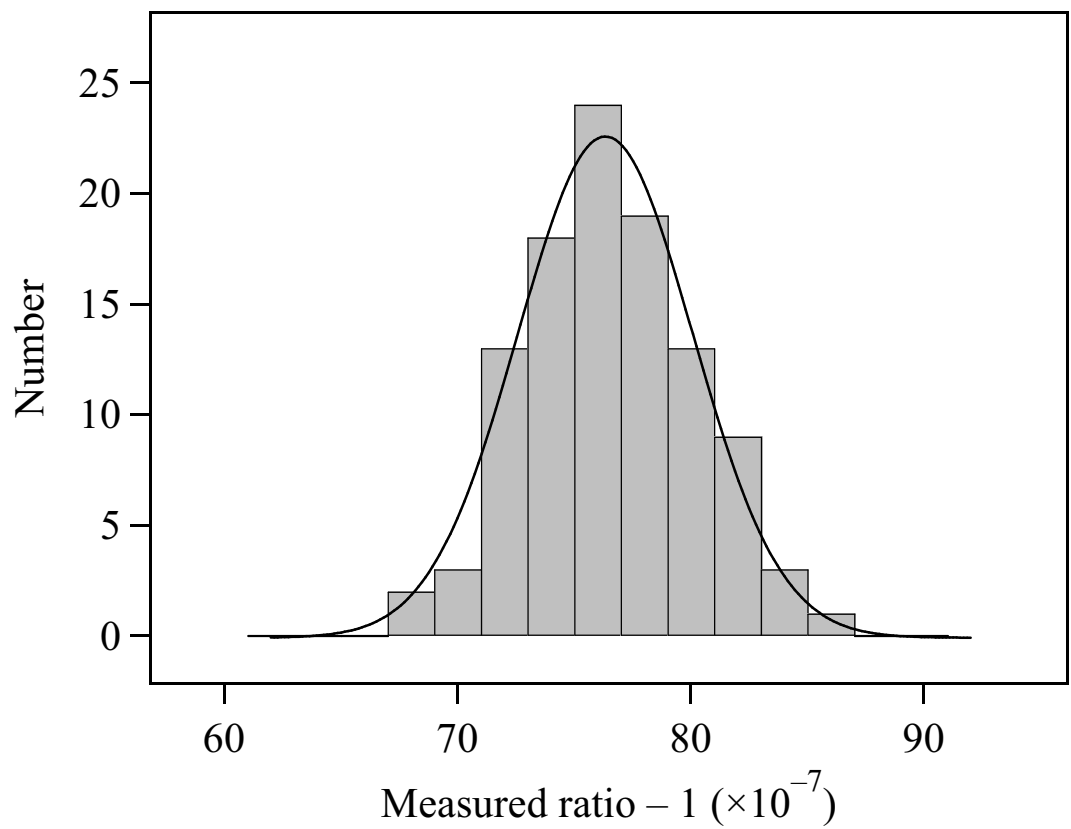
\end{document}